# Chiral-induced switching of antiferromagnet spins in a confined nanowire

T. H. Kim[1], S. H. Han[2], and B. K. Cho[1,†]

[1]*School of Materials Science and Engineering, Gwangju Institute of Science and Technology (GIST), Gwangju 61005, Republic of Korea*

[2]*Division of Navigation Science, Mokpo Maritime National University, Mokpo 58628, Republic of Korea*

[†] Correspondence and requests for materials should be addressed to B. K. Cho. (chobk@gist.ac.kr).

In the development of spin-based electronic devices, a particular challenge is the manipulation of the magnetic state with high speed and low power consumption. Although research has focused on the current-induced spin-orbit torque based on strong spin-orbit coupling, the charge-based and the torque-driven devices have fundamental limitations: Joule heating, phase mismatching and overshooting. In this work, we investigate numerically and theoretically alternative switching scenario of antiferromagnetic insulator in one-dimensional confined nanowire sandwiched with two electrodes. As the electric field could break inversion symmetry and induce Dzyaloshinskii-Moriya interaction and pseudo-dipole anisotropy, the resulting spiral texture takes symmetric or antisymmetric configuration due to additional coupling with the crystalline anisotropy. Therefore, by competing two spiral states, we show that the magnetization reversal of antiferromagnets is realized, which is valid in ferromagnetic counterpart. Our finding provides promising opportunities to realize the rapid and energy-efficient electrical manipulation of magnetization for future spin-based electronic devices.

# Introduction

In the development of highly efficient spintronic devices, one emerging issue is to discover and exploit novel phenomena with strong spin-orbit coupling (SOC) [1-4]. Due to scientific and technological interest, intensive research has focused on current-driven spin-orbit torques (SOT) for manipulation of magnetization. Most of experimental and theoretical works on SOT switching have been performed in a magnetic multilayer consisting of an ultrathin ferromagnets (FMs) or antiferromagnets (AFMs) and heavy metal layers [5-20]. Because SOT devices use a current, charge scattering and corresponding Joule heating inevitably occur [21]. This intrinsic property is an obstacle in reducing the switching power, although SOT efficiency is significantly improved in nanoscale devices [7, 10, 22-26]. In heterostructures, especially with structural inversion asymmetry, Dzyaloshinskii-Moriya (DM) interaction, which is also induced by spin-orbit coupling, has received attention in spin dynamics research. In the presence of DM interaction, the competition between exchange and DM interaction allows for a nontrivial topological spin configuration to exist as a ground state [27-30], i.e., spiral configuration and skymion in a confined geometry. Topological robustness has been exploited to enhance the performance of SOT devices, such as DM interaction-stabilized Néel domain wall motion [31-34] and DM interaction-assisted current-driven switching [35]. The DM interaction plays a secondary role in current-driven dynamics. However, it is rarely studied as a driving source to replace a current to initiate spin motion. Actually, a few studies performed on electric field-induced DM interactions found that the conversion efficiency is proportional to the spin-orbit coupling strength as in SOT [36-38].

Here, we report an electric field-induced magnetization switching scenario through potential barrier modulation in a nanowire, instead of the spin current. This switching is realized by changing the ground spiral state and relaxing it into a switched configuration by controlling the DM interactions. This switching scenario is different from the precessional switching mechanism driven by external torques, efficiency of which relies upon the timing of the torque and magnetization precession.

# Results

Figure 1 shows the spiral structure of antiferromagnetic insulator (AFI). Here, AFMs are aligned along the *z* axis and sandwiched by two electrodes of heavy and normal metal. We use two order parameters: the Néel order $\mathbf{l} = (\mathbf{s}_i - \mathbf{s}_j)/2$ and the ferromagnetic order parameter $\mathbf{m} = (\mathbf{s}_i + \mathbf{s}_j)/2$, where each spin is normalized by its magnitude $\mathbf{s}_i = \mathbf{S}_i/S_0$ with $S_0 = |\mathbf{S}_i|$. Therefore, the wire length is defined as $l_w$ in Néel space. In heavy metal layer with strong SOC such as Pt, Ta and W, spin Hall current is typically generated when a charge current is applied in those materials [1, 2]. The magnetic crystalline anisotropy has an easy axis along the *z* axis where anisotropy constant $K_z$ is positive. The geometric inversion asymmetry induces DM interaction along the *y* axis according to $\mathbf{D}_{ij} \propto \hat{\mathbf{x}} \times \mathbf{e}_{ij}$, where the *x* axis is normal to the interface and $\mathbf{e}_{ij}$ is the unit vector connecting neighbor spins $\mathbf{s}_i$ and $\mathbf{s}_j$ [27, 28]. We ignore this geometric DM interaction in the calculation and discuss it later. When the electric field along the *x* axis breaks inversion symmetry, the DM vector $\mathbf{D}$, becomes effectively toward the *y* axis due to $\mathbf{D}_{ij} \propto E\hat{\mathbf{x}} \times \mathbf{e}_{ij}$ [27, 28, 36-38]. Also, we introduce an electric-field-induced pseudo-dipolar anisotropy energy $K_E$ with easy plane, and it is induced from SOC that gives rise to the DM interaction [27, 28, 36-38]. An electric-field-induced anisotropy is an effect of order of $E^2$, but cannot be ignored in our switching scenario. In other indirect exchange interactions, known as double-exchange and Ruderman-Kittel-Kasuya-Yosida (RKKY) interaction in metal, it has been reported that same SOC induces the DM interaction and the anisotropy by the external electric field [39-41].

**Two possible spiral states as a function of DM interaction**

However, before preceding to the electric-field-induced manipulation of AFMs, we consider stationary states of AFMs as a function of DM interaction energy. Figure 2 shows two spiral structures with different DM interaction energies and these are formed by additional coupling with crystalline anisotropy, which is proportional to $\sim l_z^2$ (see Eq. (1)). Under exchange approximation where the exchange energy *J* is the larger than other energies, or $|J| \gg D_y$ and $K_z$, we can assume that the spiral structure has continuously varying spin texture, $(\mathbf{l}_{i+1} - \mathbf{l}_i)/\Delta \sim \mathbf{l}' = d\mathbf{l}/dz$ and $(\mathbf{m}_{i+1} - \mathbf{m}_i)/\Delta \sim \mathbf{m}' = d\mathbf{m}/dz$, where $\Delta$ is the interspacing of the nearest neighbors in Néel space. Therefore, the energy density $E_{1D}$ is described as

$$E_{1D} = a/2|\mathbf{m}|^2 + A/2|\mathbf{l}'|^2 + L(\mathbf{m}\cdot\mathbf{l}' - \mathbf{l}\cdot\mathbf{m}') - K_z/2(\mathbf{l}\cdot\hat{\mathbf{z}})^2 + D_y\hat{\mathbf{y}}\cdot(\mathbf{l}\times\mathbf{l}'). \quad (1),$$

The $a$ and $A$ are the homogeneous and inhomogeneous exchange constants, respectively, and $L$ is a parity-breaking exchange constant. Equation (1) is obtained in refs. 42 and 43, and the DM energy is derived in the Supplementary Note 1. The parameters are defined as $A = \Delta^2 J = J$, $a = 4J$, $L = \Delta J = J$, $D_y = \Delta d_y/2 = d_y/2$ when $\Delta$ is set to the unit length. To estimate the equilibrium state of AFMs, we set the effective Néel vector as $\mathbf{l} = \{l_x, l_z\} = \{\sin[\varphi(z)], \cos[\varphi(z)]\}$ because $l_y$ is the spiral axis. Therefore, $E_{1D}$ is reduced as

$$E_{1D} = (-\cos[\varphi]^2 K_z + d_y \varphi' + \Omega \varphi'^2)/2, \quad (2)$$

where $\Omega \equiv (-A + 3L^2/a) = J/4$ is defined as the effective exchange stiffness. After we use a standard variation of calculus to minimize total energy, $E_{total,1D} = \int_1^{l_w} E_{1D} dz$, we obtained two equations

$$\cos[\varphi]\sin[\varphi]/\Lambda^2 + \varphi'' = 0, \quad (3a)$$

$$\left|\frac{\partial \varphi'}{\partial z}\right|_{z=1 \text{ or } l_w} = \frac{d_y}{2\Omega}, \quad (3b)$$

where $\Lambda \equiv \sqrt{\Omega/K_z}$ is the characteristic antiferromagnetic domain wall width. Equation (3a) shows the stationary configuration of AFMs and takes the form of a time-independent sine-Gordan (SG) model [44]. The solution of Eq. (3a) is given as a the trivial solution $\varphi = 0$ or the nontrivial solution $\varphi = \text{am}(u|m)$. The nontrivial solution of SG equation is analytically obtained as $\varphi(z) = \text{am}[\sqrt{1 + \Lambda^2 C_1}/\Lambda(z + C_2)|1/(1 + \Lambda^2 C_1)]$, where $\text{am}(u|m)$ is a Jacobi amplitude function with the elliptic modulus $m$ and the elliptic integral of the first kind $u$. Especially, $u$ is regarded as arc length of the unit ellipse, defined as $u = F(\varphi, m) = \int_0^\varphi r(\theta, m)d\theta$ where r is radius of the unit ellipse. And the inverse function of $u$ is Jacobi amplitude: $\varphi(z) = F^{-1}(u, m) = \text{am}(u|m)$. Therefore, in general ($m \neq 0$), $\varphi(z)$ is the nonlinear function whereas $\varphi(z)$ is the linear function if it is defined with reference to a circle ($m = 0$). Here, $C_2$ is related to the phase shift along $z$ axis and $C_1$ modulates $u$ and $m$, respectively. With exchange interaction and anisotropy fixed, DM interaction modulates the reference from a circle to an ellipse. For example, when DM interaction that prefers to be spiral dominates the effective anisotropy that prefers to be uniform, $m$ approaches to zero. Therefore, $\mathbf{l} = \{\cos[\varphi(z)], 0, \sin[\varphi(z)]\}$ is described as the trigonometric function where $\varphi(z)$ is linearly proportional to $z$; $\varphi(z) = \mathrm{am}(u|0) \sim kz$. However, when the DM interaction does not dominate the anisotropy, $m$ is nonzero, so that $\varphi(z)$ becomes the nonlinear function and $\mathbf{l}$ is defined with Jacobi elliptic function; $\cos[\varphi(z)] = \cos[\mathrm{am}(u|m)] \equiv \mathrm{cn}[u|m]$ and $\sin[\varphi(z)] = \sin[\mathrm{am}(u|m)] \equiv \mathrm{sn}[u|m]$. Also, the nontrivial states have been classified into a quasi-uniformed state and a pure spiral state by the critical DM energy, where $d_y > d_c$ changes a domain wall state into a spiral state [30]. In our system, $d_c$ is derived as $d_c = 4(\Omega K_z)^{1/2}/\pi = 2(JK_z)^{1/2}/\pi$ from inserting $\varphi(z) = -\pi/2 + 2\arctan(\exp(z+C_2)/\Lambda)$ into equation (1). However, to decrease the anisotropy energies in the confined geometry, the nontrivial states are preferred to be of symmetric (S) or antisymmetric (AS) state for $l_z$ depending on the DM energy [see Fig. 2]. Each state is characterized by the first condition that is given as $\varphi(z = l_w/2) = n\pi$ for the S state or $\varphi(z = l_w/2) = (2n+1)\pi/2$ for the AS state where $n$ is an integer. The second condition becomes Neumann-type boundary condition as Eq. (3a): $|d\varphi/dz|_{z=1 \text{ or } l_w} = d_y/(2\Omega) = 2d_y/J$. Notably, as $d_y/d_c$ is over $\sim 2$, $\varphi(z)$ approaches a linear function $\varphi(z) = \mathrm{am}[u|m] \sim kz$ with wavevector $k = 2d_y/J$. And $k$ is compatible with the edge conditions because it enters into a pure spiral regime or $m \to 0$. For example, $\varphi(z) = \mathrm{am}[0.1z\,|-0.1] \sim 0.11z$ for $d_y/d_c = 6.5$ [See Fig. 2(a), S state, red open symbols] and $\varphi(z) = \mathrm{am}[0.14z\,|-0.06] \sim 0.14z$ for $d_y/d_c = 5$ [See Fig. 2(a), AS state, blue open symbols]. Therefore, $\mathbf{l}$ is expressed as a trigonometric function; for example, $l_z = \cos[kz+n\pi]$ or $l_z = \cos[kz+(2n+1)\pi/2]$. However, when $d_y/d_c$ goes to zero, $\varphi(z)$ becomes a nonlinear Jacobi amplitude function, where $\varphi(z) = \mathrm{am}[0.02z\,|-2.47]$ for $d_y/d_c = 1$ and $\varphi(z) = \mathrm{am}[0.29(z-101/2)\,|-1.40]$ for $d_y/d_c = 1.3$, respectively, because DM energy competes with anisotropy energy.

Next, we derive the dynamics of the soliton in the pure spiral regime because it provides the information about the potential barrier between two symmetric states. To understand soliton dynamics driven by damping-like SOT, the Landau Lifshitz Gilbert (LLG) equations are derived from Eq. 1 in terms of $\mathbf{m}$ and $\mathbf{l}$:

$$\dot{\mathbf{l}} = (\omega_m - \alpha\dot{\mathbf{m}}) \times \mathbf{l} + \Pi_{SOT,l}, \qquad (4a)$$

$$\dot{\mathbf{m}} = (\omega_l - \alpha\dot{\mathbf{l}}) \times \mathbf{l} + \Pi_{SOT,m}, \qquad (4b)$$

where the effective magnetic field is obtained from the functional derivative of energy density as $\omega_m/\gamma \equiv \mathbf{h}_{eff} = -\partial U_{1D}/\partial \mathbf{m} = -a\mathbf{m} - L\mathbf{l}'$ and $\omega_l/\gamma \equiv \mathbf{h}_{eff} - \partial U_{1D}/\partial \mathbf{l} = A\mathbf{l}'' + L\mathbf{m}' + K_z l_z \hat{\mathbf{z}} + \mathbf{l}' \times \mathbf{D}$, and the damping-like SOTs for $\mathbf{m}$ and $\mathbf{l}$ are given as $\Pi_{SOT,m} = \omega_s[\mathbf{m} \times (\mathbf{m} \times \mathbf{p}) + \mathbf{l} \times (\mathbf{l} \times \mathbf{p})]$ and $\Pi_{SOT,l} = \omega_s[\mathbf{m} \times (\mathbf{l} \times \mathbf{p}) + \mathbf{l} \times (\mathbf{m} \times \mathbf{p})]$, respectively [45]. $\gamma$ is the gyromagnetic ratio, $\mathbf{p}$ is the unit polarization of the spin current, $\omega_s$ is the SOT strength with an angular frequency unit and $\alpha$ is a phenomenological damping constant. The stationary state is calculated from the LLG equation when the time goes to infinity and therefore is the solution of the SG model as shown in Fig. 2.

By taking the cross product of $\mathbf{l}$ in Eq. (4a), we obtained the analytical relation between $\mathbf{m}$ and $\mathbf{l}$:

$$\mathbf{m} = \dot{\mathbf{l}} \times \mathbf{l}/(a\gamma) - L/a(\mathbf{l} \times \mathbf{l}') \times \mathbf{l}. \qquad (5)$$

To set a trial function for $\mathbf{l}$, we introduced the collective coordinates $\theta(t)$ for the dynamic phase and $k$ for the pure spiral soliton profile: $\varphi(z, t) = k(z-(l_w+1)/2)+\theta(t)$, where we arbitrary shift the soliton profile by $(l_w+1/2)$ so that $\theta(t)$ represents the phase at center or $\varphi(z = (l_w+1)/2, t) = \theta(t)$. Inserting Eq. (5) into Eq. (4b) and integrating the sublattice number $N$ from $N = 1$ to $N = l_w$, the soliton equation of motion is derived as:

$$\ddot{\theta}(t)/(a\gamma) + \alpha\dot{\theta}(t) - \Gamma(d_y, l_w)\sin(2\theta(t))\gamma/2 = w_s p_y(t) \qquad (6)$$

Equation 6 represents the equation of motion on $\theta(t)$, driven by SOT. When SOT with spin polarization of the $y$ axis applies to the antiferromagnetic chain, the soliton phase oscillates with decay as in pendulum. When SOT is strong enough for $\theta(t)$ to go over the potential barrier, $E_{barrier}$, the Néel spiral soliton propagates as a wave in medium. Here, $E_{barrier}$ is interpreted as $\Gamma(d_y, l_w) = -K_z\mathrm{sinc}[2d_y/J(l_w-1)]$, which is calculated by integrating the third terms of Eq. (6) from $\theta = n\pi$ to $\theta = (n+1)\pi/2$ or $\int_{n\pi}^{(n+1/2)\pi} \Gamma\sin[2\theta]d\theta = \Gamma$. As shown in Fig. 3(a), this potential barrier modulation effect will be negligible with large $d_y$ (or large $l_w$) because $\Gamma(d_y, l_w)$ follows the cardinal sine or sinc function. With $\dot{\theta}(0) = 0$ and $\Gamma(d_y, l_w) = 0$, the soliton phase propagates with steady state velocity $v = w_s/\alpha$ as in domain wall motion driven by SOT. Note that in Fig. 3(a), $E_{\text{barrier}}^{\text{norm}}$ is calculated from the normalized anisotropy difference between two states in Fig. 3(b) and is comparable to $\Gamma$ in the pure spiral regime, as shown in Fig. 3(a). For example, when $d_y/d_c = 3.5$, $\Gamma < 0$, $\theta(\infty)$ would be $n\pi$ (S state), which is located at potential minimum; thus, $\theta(\infty) = (n+1/2)\pi$ corresponds to potential maximum (AS state). Therefore, the anisotropy energy difference between the S and AS states are interpreted as $E_{\text{barrier}}$. The former is enable us to calculate $E_{\text{barrier}}^{\text{norm}}$ in all ranges of DM energies, as shown by the solid line in Fig. 3(a), without deriving equations of motion in low DM energy. It is difficult to derive equations of motion for soliton dynamics in cases of low $d_y/d_c$ because the soliton configurations for the S and AS states consist of different wavepackets [see the Supplementary Note 2 for S and AS states when $d_y/d_c = 1$] and there is a deviation between the SG model and the pure spiral model [see Fig. 3(a)].

**Electric-field-induced switching of antiferromagnetic solitons.**

Now, the electric-field-induced DM interaction and easy-plane anisotropy are considered. Firstly, the anisotropy in Eq. (1) is recast into $\sum E_K = -K_{\text{eff}}/2(\mathbf{l}\cdot\hat{\mathbf{z}})^2 - K_E/2(\mathbf{l}\cdot\hat{\mathbf{y}})^2$ where $K_{\text{eff}} = K_z + K_E$. The $y$ component of the easy plane anisotropy does not contribute to the stability of spiral states because of $K_{\text{eff}} > K_E$. And we reformulate the easy plane anisotropy energy as a function of $d_y$; for example, if the DM interaction is induced by electric field as like $d_y = 0.1J$, $K_E = 0.1^2 J$ or $K_E = (d_y/J)^2 J$. Now, $d_c$ and $\Lambda$ are as a function of $d_y$; in large $d_y$, $K_{\text{eff}} \sim K_E$ and $d_c = 2(JK_E)^{1/2}/\pi = 2d_y/\pi$, and $\Lambda \equiv \sqrt{\Omega/K_{\text{eff}}} = \sqrt{J/K_{\text{eff}}}/2 = J/(2d_y) \sim 0$. It means that in all ranges of $d_y$, the pure spiral configuration is not formed, so that all soliton states are not described as $\theta(t)$ because $d_y/d_c < 1.5$. However, the S and AS states are calculated numerically using two conditions in SG model [see Figs. 4(a) and 4(b)]. The curves of the anisotropy energy and the potential barrier are not derived analytically.

To switch Néel magnetization, our strategy is to modulate potential barriers by controlling ratio $d_y/J$ through several steps in which SOT plays a perturbation role. As shown in Fig. 4(b), the stationary soliton state is alternatively changed from S to AS states ($E_{barrier} < 0$ to $E_{barrier} > 0$ in Fig. 4(a)) and then from AS to S states ($E_{barrier} > 0$ to $E_{barrier} < 0$ in Fig. 4(a)) as the DM energy increases. It completes the Néel arrangement switch in the five steps. In $d_y/J = 0$ (step 1), the uniform antiferromagnetic state along the +z axis is interpreted as an S state with $\varphi(l_w/2) = 0$. Although the DM energy turns on when $d_y/J = 0.043$ (step 2), the soliton state is not changed because $\varphi(l_w/2) = 0$ and $E_{barrier} < 0$. When the DM energy is lowered by $d_y/J = 0.03$ (step 3), the S state is unstable because $E_{barrier} > 0$ [see Fig. 4(a), AS-state] but, interestingly, does not go into the AS state because it is a metastable state located at a potential maximum [see Fig. 4(c)], which implies the necessity of small perturbation such as SOT. Therefore, small SOT with unidirectional polarization is necessary for deterministic switching. For example, with a spin current with $-p_y$, the soliton would go to an AS state with $\varphi(l_w/2) = -1/2\pi$; if spin polarization is of $p_y$, AS state would be of $\varphi(l_w/2) = 1/2\pi$. Next, in the lowered $d_y/J = 0.015$ (step 4), AS state with $\varphi(l_w/2) = -1/2\pi$ is required to go S state with $\varphi(l_w/2) = -\pi$. Eventually, as DM energy shuts down (step 5), final S state is maintained with $\theta = -\pi$. All processes are described in Fig. 5(a). Note that our solitonic approach allows for simplifying the multistep manipulation of AFMs; because the first two steps and the fourth and fifth steps are in the same state of $\varphi(l_w/2) = 0$ and $\varphi(l_w/2) = -\pi$, so these overlapping steps could be omitted. As shown in Fig. 5(b), only the first, third and fifth steps that form the single pulse shape can switch an AFM. In addition, the $d_y$ variation from step 1 to step 2 results in spreading and shrinkage of $k$, i.e., breathing motion due to inertia. However, this motion does not lead to the phase propagation. In addition, it is desirable to consider the field-like torque taking place during working in the real devices. When the magnetic field is applied along arbitrary directions, we can add the Zeeman interaction energy $E_{1D, Z} = \gamma\hbar \mathbf{H} \cdot \mathbf{m}$ into the total energy density, where $\gamma$ is the gyromagnetic ratio and $\hbar$ is the reduced Plank constant. And Eq. (5) is modified as $\mathbf{m} = \mathbf{l} \times \mathbf{l'}/(a\gamma) - L/a(\mathbf{l} \times \mathbf{l'}) \times \mathbf{l} + \gamma\hbar/a\mathbf{H}$ [43]. If the magnetic field is time-varying, the spiral soliton is driven by field-like torque, $\sim dh_y/dt$ [46], which is derived after inserting Eq. (5) into Eq. (4b). To suppress field-like torque, the proper strength of SOT should be applied.

**Discussion**

Our proposal is dependent of the proper size ($l_w = 100$ or $2N = 200$ spins in this calculation)

and electric field strength; the necessary electric field can be easily estimated. The characteristic spin-orbit coupling energy $E_{SO}$ of $Y_3Fe_5O_{12}$ garnet is 3 eV [38] and in transition metal compounds, $E_{SO}$ is typically on the order of ~ 3 eV. Therefore, the electric field, required to generate $d_y / J = 0.043$ can be estimated as $|E| = E_{SO}D/(Jed_a)$ ~ 0.13 Vnm$^{-1}$, where $d_a$ is the distance between the nearest neighbor magnetic ions and is set to be ~ 1 nm. To estimate switching power in our work, we suppose simple magnetic pad geometry with thickness $t$ (= 5 nm), width $w$ (= 60 nm) and pad length ($2l_w$ = 200 nm). In the nano-pad with finite $w$, the two DM vectors ($D_y$ and $D_z$) are induced, according to $\mathbf{D}_{ij} \propto E\hat{\mathbf{x}} \times \mathbf{e}_{ij}$, However, the effective anisotropy is along $z$ axis, so that $D_z$ induces magnon splitting in momentum space, not spiral structure along $y$ axis. Possible candiates are MnF$_2$ [47], and YFeO$_3$ [48]; all are G-type antiferromagnetic insulators with dominant easy axis and the ratio $K/J$ ~ $10^{-4}$. For example, at room temperature, the resistivity of YFeO$_3$ is $\rho$ ~ $10^6$ ohm·m [49] and the resistance $R = \rho t/(wl_w)$ ~ $4.17 \times 10^{11}$ ohm. Therefore, power $W = V^2/R$ ~ 1 pW.

Table I, we compare the critical switching power estimated from our scenario and SOT or spin transfer torque (STT) devices in the different magnetic tunnel junction structures. The different types of SOT and STT devices is characterized to be of comparable order from ~ μW to ~ mW where SOT devices have the mimimum size as determined by thermal requirements [50, 51].

Table I. Comparison of switching power between the different magnetic structures reported to date.

| | IMA AFM | IMA FM W/CoFeB/MgO | IMA FM Pt/CoFeB/MgO | PMA FM Ta/CoFeB/MgO |
|---|---|---|---|---|
| Driving force | Electric field | SOT | SOT | STT |
| Critical current density or Electric field | 0.13 Vnm$^{-1}$ | 5.4 MAcm$^{-2}$ | 40 MAcm$^{-2}$ | 4 MAcm$^{-2}$ |
| Power (W) | 1 pW | 52 μW | 41 mW | 75 μW |
| Reference | Our work | Shi et al. [50] | Aradhya et al. [51] | Ikeda et al. [52] |

Acronyms: AFM, antiferromagnet; FM, ferromagnet; SOT, spin-orbit torque; STT, spin transfer torque.

As noted in introduction, structural DM interaction strength by asymmetric electrodes could be reduced below $d_c$ by engineering its thickness [53] or utilizing symmetric electrodes, compared with electric field-induced DM energy. However, the structural DM interaction, weak enough to form a quasi-uniform configuration, reduces the required electric field strength. The above statements are also valid in ferromagnetic counterparts because a ferromagnetic spiral structure is formed by competition between anisotropy and DM energy and is excited by SOT; in ferromagnetic nanowire, two conditions are given as $\varphi(z = l_w/2) = n\pi$ for the S state or $\varphi(z = l_w/2) = (2n+1)\pi/2$ for the AS state and $d\varphi/dz|_{z=1 \text{ or } l_w} = d_y/(2A)$ and $d_c = 4(AK_{\text{eff}})^{1/2}/\pi$ [30]. Finally, it remains to be seen if there is the electric field effect in different magnetic systems. In magnetic metal system with broken inversion symmetry, the generation mechanisms of DM interaction are two folds: 1) Fert-Levy mechanism [54] and 2) Rashba SOC [39-41, 55]. In the Fert-Levy mechanism, itinerant electron is mainly exchange-coupled with magnetic ion by RKKY interaction. An additional coupling leads to the DM interaction by scattering of itinerant electron with heavy metal. As aforementioned, the Rashba SOC is related to also itinerant electron in the material with strong SOC. Another electric field induced modulation of anisotropy is reported in the ferromagnetic metal/oxide interface or Ta/ultrathin CoFeB/MgO [56], the perpendicular magnetic anisotropy (PMA) is originated from hybridization of oxygen *p*-orbital and iron *d*-orbital. In this case, the electric field induces charge redistribution of electron of magnetic metal, resulting in modulation of PMA [21, 56, 57]. However, the magnetic insulator is lack of conduction electron and it is hard to expect the charge redistribution by electric field and its related anisotropy modulation.

In conclusion, we investigated spiral dynamics in the presence of DM interaction. In soliton-based spin dynamics, there are two states (symmetric and antisymmetric state) due to competition between anisotropy energy and DM interaction, in which one is stable at a potential minimum, and the other is metastable at a potential maximum, implying that external (or internal) perturbation is necessary for viable applications. Also, all points with potential barrier of zero should be avoided because a single state is not determined energetically. Electric field control of DM interaction is promising for manipulation of AFM because it overcomes the challenging issues of phase matching and overshooting by conventional external torque and does not induce charge-carrying issues such as Joule heating. Finally, by tuning the DM energy and interpreting spiral behavior on soliton picture, we show that the AFM switching can be performed with an effective single-pulse scheme.

# Method

### Numerical simulations

Numerical simulations (Landau Lifshitz Gilbert model, equations (4a) and (4b)) were conducted from 0 to 500 picosecond (ps) with a 0.1 ps interval using proper parameters for antiferromagnetic insulators with terahertz precessional frequency: $J = 41.4$ meV ($10^{13}$ s$^{-1}$, 10 THz), $K_z = 0.0003J$ or 4.14 μeV ($10^9$ s$^{-1}$, 1 GHz), $\omega_s = 2\pi \times 10^4$ s$^{-1}$ ($<< K_z$), $\alpha = 0.0008$ and $l_w = 100$. The magneto-static interaction is neglected for clear oscillating behavior of phase. The rising and falling times of the time-varying electric field pulse were set to 1 ps so that the oscillating phase does not experience unwanted effects during electric field change.

# DATA availability

The data that supports the findings of this study is available from the corresponding author upon request.

## Acknowledgements

This work was supported by GIST Research Institute (GRI) a grant funded by the GIST in 2019 and by National Research Foundation of Korea (NRF), funded by the Ministry of Science, ICT & Future Planning (No. NRF-2015M3A9B8032703, No. NRF-2017R1A2B2008538 and NRF-2018R1A2B6005183)

## Author Contributions

B.K.C. and T.H.K. conceived the project idea and planned the analytical and numerical calculations. T.H.K. performed the analytical and numerical calculations. T.H.K., S.H.H., and B.K.C. analyzed the data. B.K.C. led the work and wrote the manuscript with T.H.K. The results of the theoretical and numerical findings were discussed by all coauthors.

## Additional information

**Competing interests:** The authors declare no competing interests.

# Figure Legends

**Fig. 1** Schematics for antiferromagnetic spiral structure in Néel space. In a confined one-dimensional geometry, this spiral structure is formed by Dzyaloshinskii-Moriya (DM) interaction. When electric field is applied between two electrodes and DM interaction is induced, the DM vector between neighboring spins takes the form of $\mathbf{D}_{ij} \propto \hat{\mathbf{x}} \times \mathbf{e}_{ij}$, where $\mathbf{e}_{ij}$ is the unit vector linking neighbor spins $i$ and $j$ and therefore, $\mathbf{D}_{ij}$ is parallel to the $y$ axis. Here, spin-orbit torque with polarization along the $y$ axis is applied to perturb the antiferromagnets.

**Fig. 2.** Equilibrium configurations of antiferromagnets in a confined structure. Depending on the Dzyaloshinskii-Moriya (DM) interaction, $l_z = \cos(\varphi)$ takes a (a) symmetric (S) or (b) antisymmetric (AS) configuration. Here, $\varphi$ is described as a Jacobi amplitude function $\varphi = \mathrm{am}(u|m)$ with elliptic modulus $m$ that is the solution of the sine-Gordan (SG) equation. The exact solution is obtained with two conditions: 1) $\varphi(l_w/2) = \pi$ or 0 and 2) $d\varphi/dz = 2d_y/J$ at $z = 1$ or $z = l_w$ where $l_w$ is wire length in Néel space. The stationary state is calculated from the Landau Lifshitz Gilbert equation when the time goes to infinity and therefore is the solution of the SG model (solid line). As the DM energy increases, $l_z$ becomes a pure spiral configuration of $\varphi = kz$ with wavevector $k = 2d_y/J$ (solid line and open circle for $n = 5$). However, as the DM energy decreases, there is a deviation between the pure spiral approximation and the SG model (solid line and open circle for $n = 2$). Here, $d_c$ is the critical DM energy where $d_y > d_c$ changes a domain wall state into a chiral state.

**Fig. 3.** Two possible spiral states as a function of Dzyaloshinskii-Moriya (DM) interaction. (a) The potential barrier that is calculated from $E_{\mathrm{barrier}}^{\mathrm{norm}} = E_{\mathrm{ani}}^{\mathrm{S, norm}} - E_{\mathrm{ani}}^{\mathrm{AS, norm}}$ in the range of DM energy. Note that when $d_y/d_c \gg 2$, so that the soliton is well-described by the pure spiral configuration $l_z = \cos(kz)$, $E_{\mathrm{barrier}}^{\mathrm{norm}}$ is described as a cardinal sine or sinc function $\Gamma_{\mathrm{norm}}$, implying that $E_{\mathrm{barrier}}^{\mathrm{norm}}$ is negligible with large DM interaction or the long length $l_w$. (b) The normalized anisotropy energies, $E_{\text{ani}}^{\text{S, norm}}$ (symmetric state, S state) and $E_{\text{ani}}^{\text{AS, norm}}$ (antisymmetric state, AS state).

**Fig. 4.** Two possible spiral states determined by Dzyaloshinskii-Moriya (DM) interaction and easy plane anisotropy. (a) The potential barrier that is calculated from $E_{\text{barrier}}^{\text{norm}} = E_{\text{ani}}^{\text{S, norm}} - E_{\text{ani}}^{\text{AS, norm}}$ in the range of DM energy. Note that $d_c$ are as a function of $d_y$; in large $d_y$, $K_{\text{eff}} \sim K_E$ and $d_c = 2(JK_E)^{1/2}/\pi = 2d_y/\pi$ or $d_y/d_c \sim 1.5 < 2$, so that the soliton is not described by the trigonometric function $l_z = \cos(kz)$ (b) The normalized anisotropy energies, $E_{\text{ani}}^{\text{S, norm}}$ (symmetric state, S state) and $E_{\text{ani}}^{\text{AS, norm}}$ (antisymmetric state, AS state). (c) Schematics for potential depending on the sign of $E_{\text{barrier}}^{\text{norm}}$. For example, when $E_{\text{barrier}}^{\text{norm}} < 0$ (or $E_{\text{barrier}}^{\text{norm}} > 0$), the S state (or AS state) is energetically stable with minimum potential energy and the AS state (or S state) is metastable with maximum potential energy.

**Fig. 5.** Antiferromagnetic switching through Dzyaloshinskii-Moriya (DM) energy changes. DM pulse with (a) multistep or (b) single-step profile is applied to induce antiferromagnetic switching, applying weak spin-orbit torque (SOT). Here, electric-field-induced anisotropy is as a function of DM energy. According to potential barrier profiles, the first two steps ($d_y/J = 0$ and 0.043) are stable with a symmetric (S) state. At $d_y/J = 0.03$, the S state is metastable and antisymmetric (AS) becomes a stable state with $\varphi(l_w/2) = -1/2\pi$ (not $1/2\pi$) due to unidirectional SOT with spin polarization $-p_y$. In the fourth and fifth steps ($d_y/J = 0.015$ and 0), the AS state is metastable. Thus, the S state has $\varphi(l_w/2) = -\pi$ (not $\pi$) due to $-p_y$. In our switching scenario, the first two steps and the fourth and fifth steps overlap. Therefore, the Néel order could be switched using a single-step function without the second and fourth processes. (c) Schematic illustration of the Néel configuration for each step.

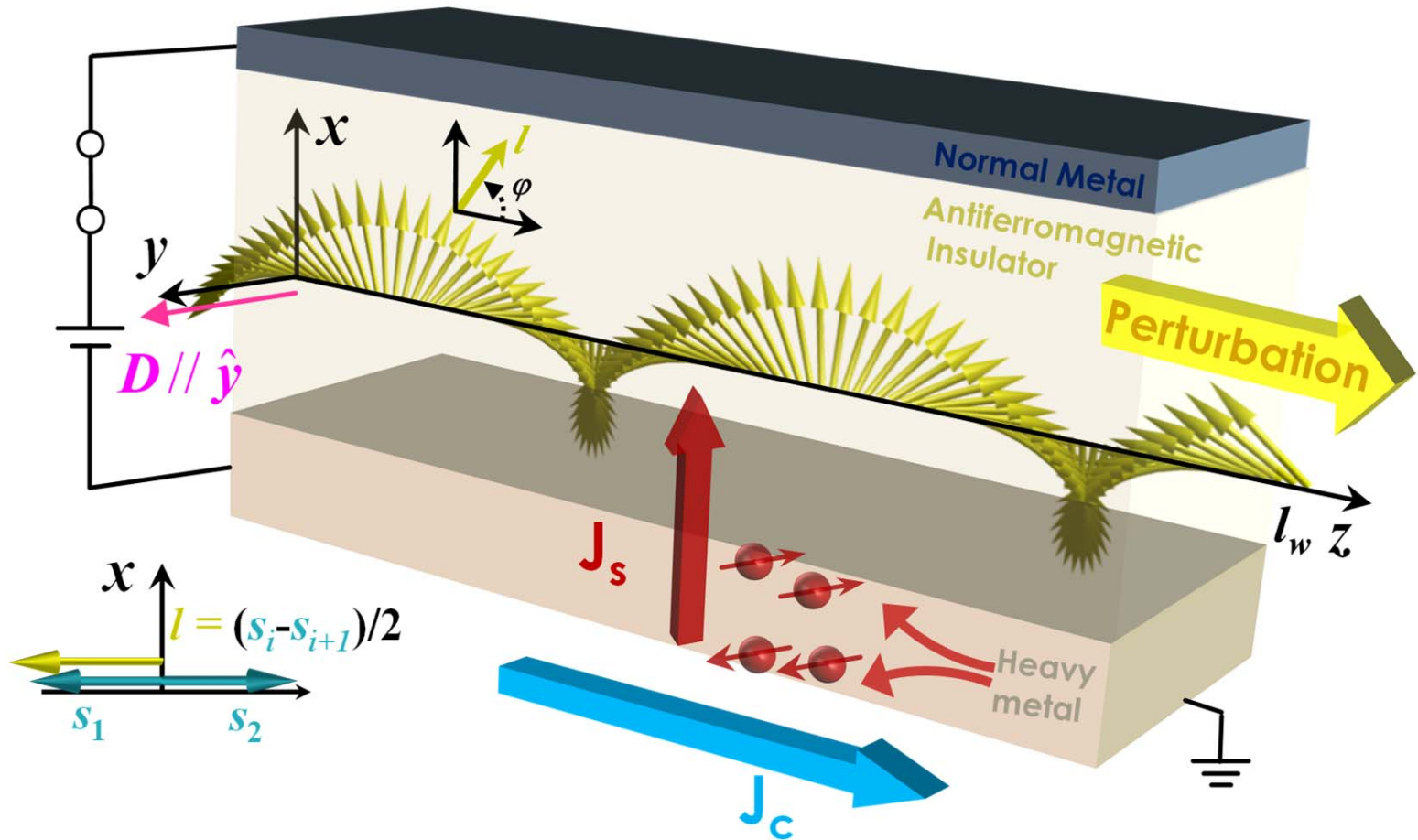

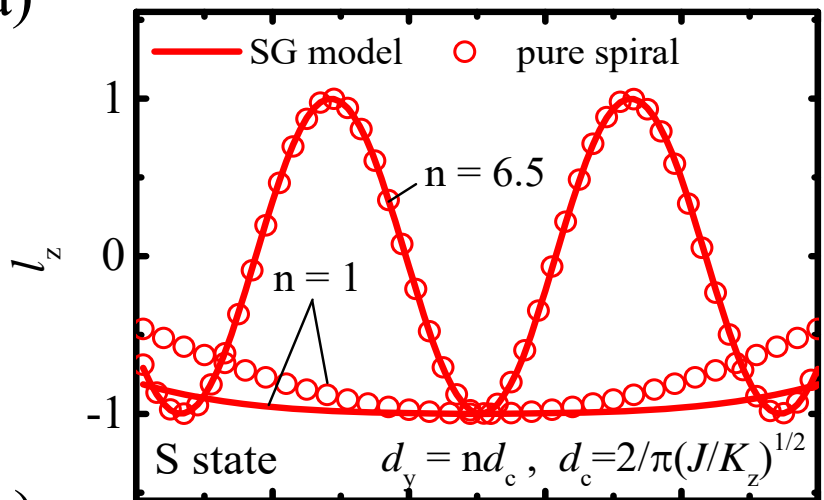
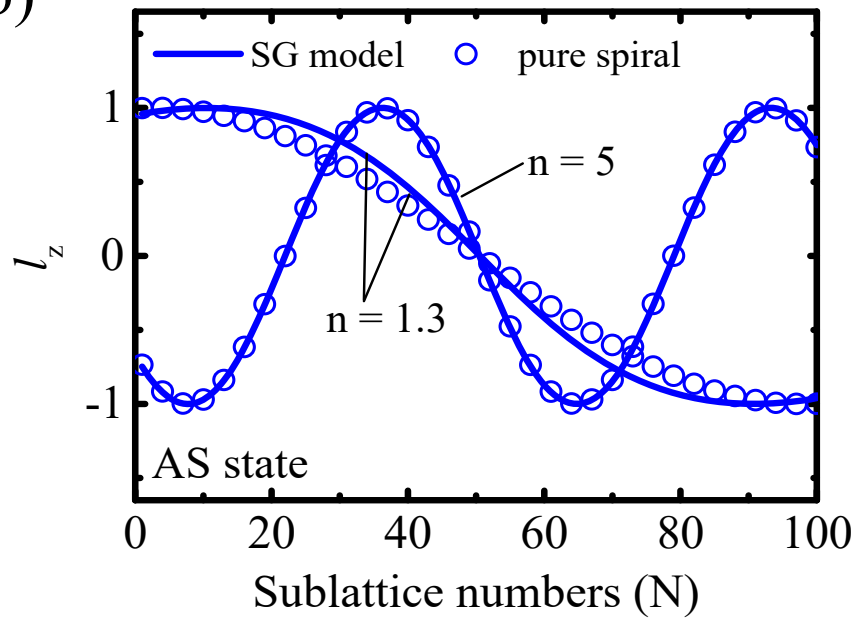

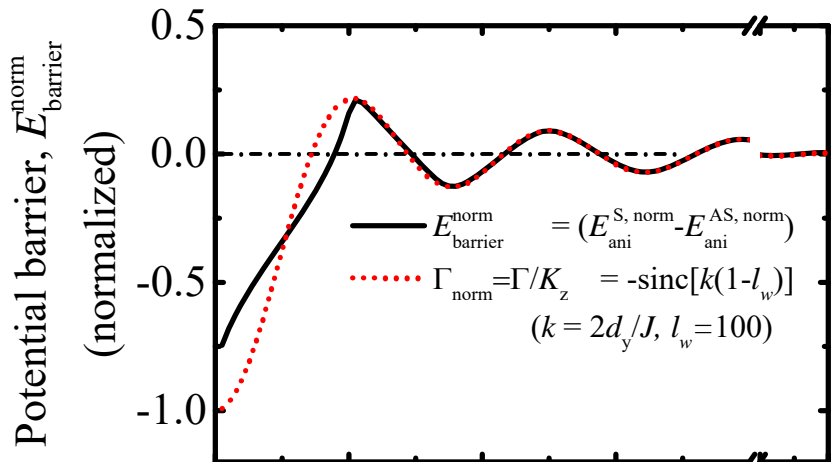
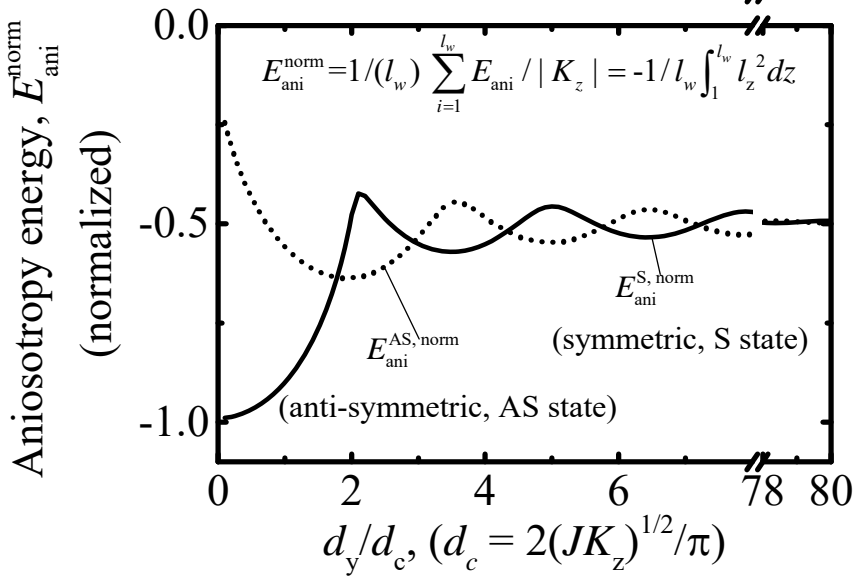

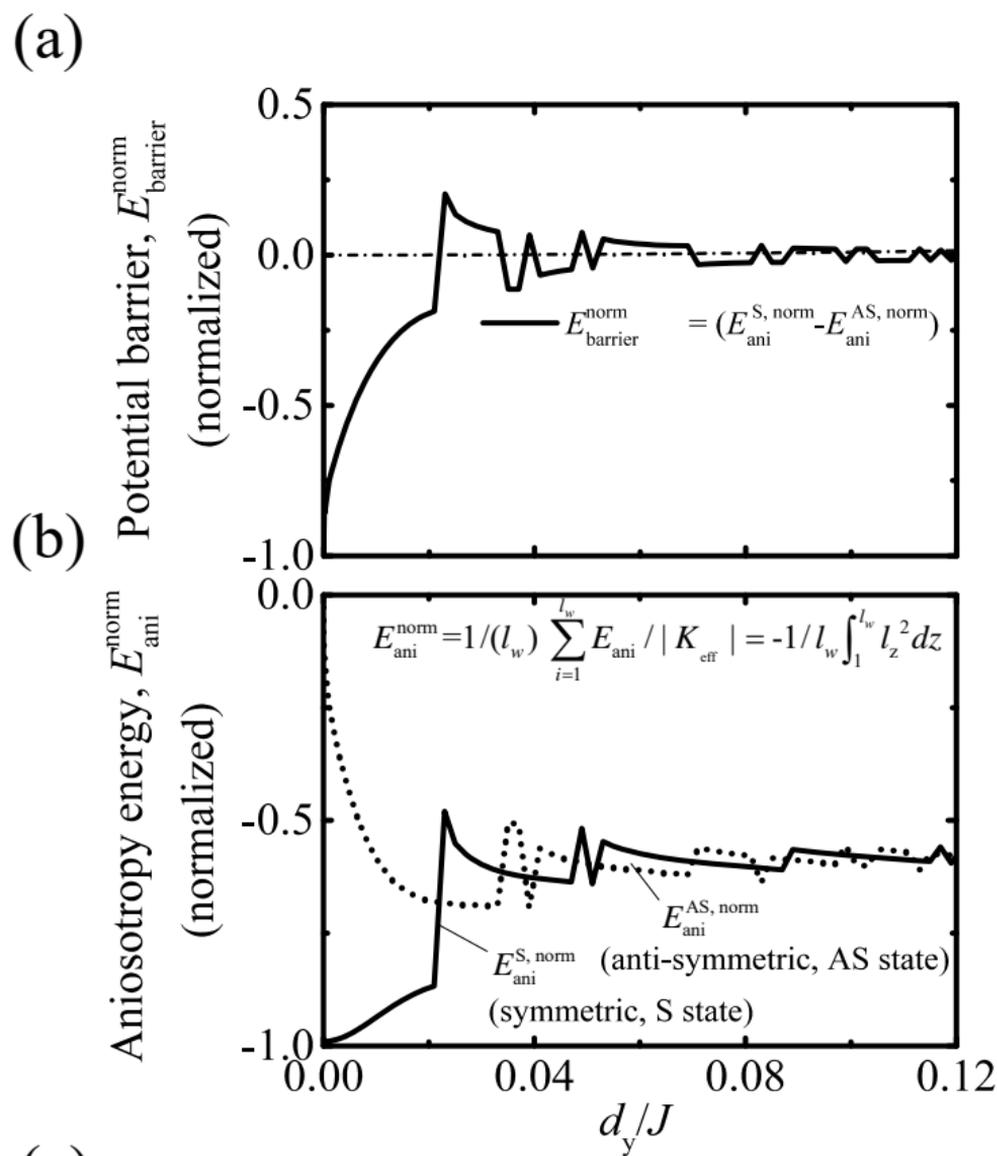

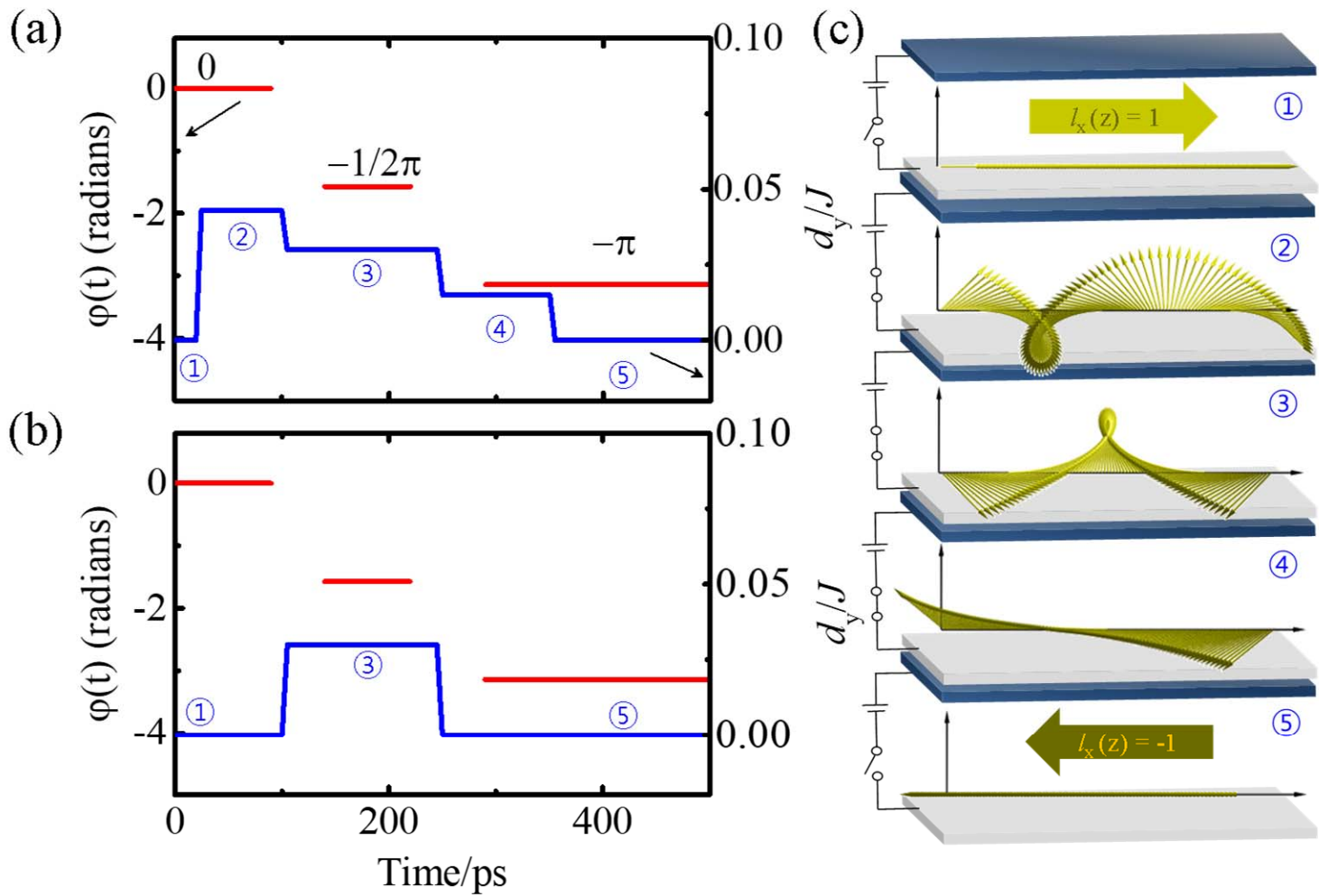